\documentclass[aps,10pt,prl,twocolumn]{revtex4-2}

\usepackage{graphicx}
\usepackage{comment}

\usepackage{graphicx}
\usepackage{epstopdf}
\usepackage{dcolumn}
\usepackage{amsmath,amssymb,dsfont}
\usepackage{wrapfig}
\usepackage{array}
\usepackage[caption=false]{subfig}
\usepackage{tabularx}
\usepackage{array}
\usepackage{multirow}
\usepackage{booktabs}

\renewcommand{\sc}{n_{\mathrm{{\scriptscriptstyle 2D}, int}}}
\newcommand{\scu}{10^{12}/\text{cm}^2}

\usepackage{listings}
\usepackage[bookmarks=false,linkcolor=blue,urlcolor=blue,colorlinks,citecolor=blue]{hyperref}
\usepackage{exscale,relsize}
\usepackage[usenames, dvipsnames]{color}
\usepackage{bbold}
\usepackage{xcolor}

\usepackage{cleveref}
\Crefformat{appendix}{Appendix~#2#1#3}

\usepackage{microtype}

\usepackage{siunitx}
\DeclareSIUnit{\ueV}{\micro\electronvolt}
\DeclareSIUnit{\um}{\micro\meter}
\DeclareSIUnit{\nm}{\nano\meter}
\usepackage[]{cleveref}
\Crefname{section}{Sec.}{Secs.}
\Crefname{subsection}{Sec.}{Secs.}
\Crefname{appendix}{\IfAppendix{Sec.}{Sec.}}{\IfAppendix{Secs.}{Secs.}}
\Crefname{subappendix}{\IfAppendix{Sec.}{Sec.}}{\IfAppendix{Secs.}{Secs.}}
\Crefname{equation}{Eq.}{Eqs.}
\Crefname{figure}{Fig.}{Figs.}
\Crefname{tabular}{Tab.}{Tabs.}

\newcommand{\RefTGPe}{Ref.~\onlinecite{Aghaee23}}
\newcommand{\RefTGPt}{Ref.~\onlinecite{Pikulin21}}
\newcommand{\RefLegga}{Ref.~\onlinecite{Legg25a}}
\newcommand{\RefLeggb}{Ref.~\onlinecite{Legg25b}}
\newcommand{\RefZMPR}{Ref.~\onlinecite{Aghaee25}}

\newcommand{\Bpara}{B_\parallel}

\newcommand{\Vp}{V_\mathrm{p}}
\newcommand{\Vc}{V_\mathrm{c}}
\newcommand{\Vbias}{V_\mathrm{bias}}

\newcommand{\Gth}{G_\mathrm{th}}
\newcommand{\GNL}{G_\mathrm{NL}}

\begin{document}

\title{Response to recent comments on
Phys. Rev. B 107, 245423 (2023) and\\
Subsection S4.3 of the
Supp. Info. for Nature 638, 651-655 (2025)}
\author{Microsoft Quantum$^\dagger$}
\noaffiliation
\date{\today}

\begin{abstract}
The topological gap protocol (TGP) is a statistical test designed to identify a topological phase with high confidence and without human bias. It is used to determine a promising parameter regime for operating topological qubits \cite{Aghaee23, Aghaee25}. The protocol's key metric is the probability of incorrectly identifying a trivial region as topological, referred to as the false discovery rate (FDR). Two recent manuscripts \cite{Legg25a,Legg25b} engage with the topological gap protocol
and its use in
\RefTGPe~and Subsection S4.3 of the
Supplementary Information for
\RefZMPR, although they do not
explicitly dispute the main results
of either one. We demonstrate that
the objections in Refs.~ \onlinecite{Legg25a,Legg25b} are unfounded, and we uphold the conclusions of Refs.~\onlinecite{Aghaee23, Aghaee25}. Specifically, we show that no flaws have been identified in our estimate of the false discovery rate (FDR). We provide a point-by-point rebuttal of the
comments in Refs.~\onlinecite{Legg25a,Legg25b}.
\end{abstract}

\maketitle

\begin{widetext}
    
\section{Summary of Point-by-Point Responses}

\begin{enumerate}

\item {\it Identification of the ‘gap’ differs between publication and released code} \cite{Legg25a}. This is incorrect. There is no difference \cite{Aghaee23}. In addition, it is worth emphasizing that the transport gap requires a careful definition in a finite-sized disordered system, as discussed at length in Ref.~\onlinecite{Aghaee23},
but seemingly overlooked
in \RefLegga, which uses quotation marks around the word `gap.' 

\item {\it The TGP applied to experiments is not the same TGP applied to theoretical simulations} \cite{Legg25a}. The difference \cite{Aghaee23} is very minor and is easily changed by modifying a single parameter in one line of code, which shows that there is no statistically significant difference between the two versions (1 false positive out of approximately 700 regions of interest). Minor differences between simulations and experiments pose no issues as long as the simulations enable an accurate bound on the FDR.

\item {\it Large unexplained variations in experimental data parameters that change TGP outcome} \cite{Legg25a}. The variations in experimental data parameters {\em are} explained
\cite{Aghaee23}, and TGP outcomes are not the consequence of measurement choices. Moreover, as we show below, minor variations in experimental parameters do not significantly affect the false discovery rate of the TGP.
It is not surprising that the TGP has some sensitivity to data ranges, as pointed out in \RefTGPe. 
A larger magnetic field range enables better sampling, which reduces the error bars for the transport gap. On the other hand, it is impractical to measure over too large a range, so we
strike a balance that keeps
the FDR low: we use
a Stage 1 scan to determine regions where a more detailed Stage 2 measurement is performed.

\item {\it A redefinition of ‘topological’ enables the claim of zero false positives} \cite{Legg25a}. There is no redefinition of topological within \RefTGPe, which is self-contained and has a clearly stated definition. There is an important question -- what is the best way to define topological order in a finite-sized system with disorder? – but this is not analyzed in the comment \cite{Legg25a}. The definition that we used in our paper is quite stringent. Approximately 10\% of phase space satisfies our definition, not 37\%, as claimed in the comment. Of the fraction of phase space that passes Stage 1 of the TGP, 37\% is topological, but this just shows that Stage 1 succeeds in eliminating uninteresting parts of phase space. Moreover, \RefTGPe~doesn’t claim that false positives cannot occur.

\item {\it The TGP can report the region investigated for ‘parity readout’ as both gapped and gapless} \cite{Legg25b}. This claim is based on
point 3 above, according to which
variations in experimental parameter ranges determine whether a region is classified as gapped or gapless. This is incorrect and was addressed in our response to point 3 above.

\item {\it TGP outcomes not correct, inhibiting exploration of reproducibility} \cite{Legg25b}. This is incorrect.
Reproducibility was shown by performing
two different measurements on one device and a measurement on a second device \cite{Aghaee25}. It is true that parity
measurements were not performed at
every point in these devices' phase space since such
an exhaustive survey of phase space
wasn't the goal of \RefZMPR.

\item {\it Conductance data that was not presented shows high levels of disorder and no clear superconducting gap} \cite{Legg25b}.
This is incorrect. Our conductance data, as analyzed via the TGP
\cite{Aghaee23} shows that there is a gap \cite{Aghaee25}, and the quantum capacitance measurements of the fermion parity in \RefZMPR~are not consistent with a gapless system.
Moreover, our process yields devices with low disorder, as we have shown by direct measurements of the localization
length \cite{Aghaee23}.

\end{enumerate}

\end{widetext}

\section{Introduction}
\label{sec:intro}

The topological gap protocol (TGP)
is a statistical test for
analyzing three-terminal transport
data to determine if certain features
(zero-bias peaks, gap closing/re-openings)
are sufficiently robust as to indicate
the presence of topological superconductivity
\cite{Pikulin21,Aghaee23}.
It has recently been used to
develop a material and device platform
for topological qubits based on
InAs-Al hybrid devices \cite{Aghaee23,Aghaee25,Aasen25}.
In comments on Refs.~\onlinecite{Aghaee23,Aghaee25},
two recent manuscripts 
\cite{Legg25a,Legg25b} have criticized various aspects of the topological gap protocol. In this note,
we address these criticisms~\cite{Legg25a,Legg25b}.
We believe that they
 are based on a misconception about the purpose of the topological gap protocol: it is a tuning tool to bring a qubit device into a topological regime, based on a statistical model. As such, there is a distinction between the protocol output being sensitive to the protocol's defining parameters, its ``hyperparameters,'' (which it is, as is natural for a binary test that depends on analog data) vs. frequently
incorrectly classifying trivial regions as topological (which it does not). Different realizations of the protocol with slightly different hyperparameters may well lead us to slightly different topological regions in parameter space, but they will, with high probability, not lead us to a trivial one.

We start by summarizing the reasoning behind the topological gap protocol and provide technical details relevant for the discussion here. 

A 1D topological superconductor (1DTS) is characterized by a bulk topological gap with two separated Majorana zero-energy modes (MZMs) at its ends that form a single fermionic state~\cite{Kitaev01, Lutchyn10, Oreg10}.
Many attempts at experimentally identifying a bulk topological phase have failed because there is no single measurable quantity that uniquely identifies its presence in real devices. All devices have disorder, are finite-sized, and deviate from idealized theoretical models. This leaves room for alternate explanations for the observed phenomena.
These challenges necessitate indirect methods that rely on multiple measurable quantities and their systematic dependence on experimental control parameters. It is the entirety of the data set that confines different scenarios and allows one to distinguish between them.

In \RefTGPt, we proposed a protocol to identify a transition to a topological phase in three-terminal devices by measuring the dependence of the local and non-local conductances~\cite{Rosdahl18} on  experimentally controlled parameters: terminal bias voltages, magnetic field, and voltages on electrostatic gates that tune chemical potential and junction transparencies. 
The key criteria that need to be satisfied are: the simultaneous occurrence of robust local zero-bias peaks (ZBPs) and a bulk transport gap in non-local conductance that closes and reopens as the field and the gate that tunes the chemical potential are varied, mapping out the phase diagram of the device.
While these signatures yield textbook phase diagrams in the theory of clean and sufficiently long devices, in real finite-sized devices with disorder and other imperfections, the phase diagram looks much less intuitive \cite{Stanescu11, Adagideli14, Pan21a}. The TGP offers an objective way of identifying prospective regions of interest in the measured phase diagram of the device that have a high probability of overlap with the topological phase.
It relies on filtering, thresholding, and clustering the data that was measured and requires appropriate thresholds to be set by calibrating to simulated data.
The TGP provides a binary answer with statistical confidence, determined by testing against models that capture the key characteristics of the device as verified by independent measurements. This process is explained in detail in \RefTGPe.

To optimize the TGP, we introduced the false discovery rate (FDR), which represents the probability that the protocol misidentified the region of interest, i.e., a trivial region was identified as topological. 
By using well-informed input parameters and testing the TGP against device simulations, we have demonstrated that the FDR can be reduced to a small value (below $8\%$) \cite{Aghaee23}. 
Therefore, passing the topological gap protocol implies that we have identified a topological region with high confidence. 
Failing the topological gap protocol does not necessarily mean that
the device does not
have a topological phase (i.e. false negatives are expected to occur).
Finding every possible region of topological phase in the given measurement is not a goal of the TGP.
In summary, false positives of the TGP may occur, but the probability of their occurrence is small. False negatives can occur as well but we have not quantified their probability.

The preceding summary outlines the logic behind the TGP and provides the context for the main claim of \RefTGPe, which is
that we have fabricated and measured 
InAs-Al hybrid devices that have passed the topological gap protocol.
This claim is not contested by the comment in \RefLegga.
In \RefLegga, the author argues that the
TGP is not a binary pass/fail criterion.
This statement is incorrect.
In fact, the TGP does provide a binary answer, always.
It appears that \RefLegga~
is actually challenging the probability that this answer is correct. We refute this below.

\RefLeggb~attempts to criticize
Subsection S4.3 of the online Supplementary Information for \RefZMPR.
The points raised do not provide any evidence that the FDR of the TGP is higher than claimed. Therefore, there is no reason to doubt the validity of the tuneup procedure used for the devices reported in \RefZMPR.

In the detailed technical response that follows, we explain why each specific
point raised in Refs.~\onlinecite{Legg25a,Legg25b} fails to warrant any revision of the claims
in Refs.~\onlinecite{Aghaee23,Aghaee25}.

\section{Technical response to Ref. 3}
\label{sec:TechResp}

We now address the technical comments raised in \RefLegga~point-by-point. 

\subsection{1. Algorithm for extracting gap}
\RefLegga~claims that there is a difference between the description of how the threshold factor $\Gth$ is set
in the text of \RefTGPe~and what is implemented in code~\cite{azure-quantum-tgp}.
However, this is not the case.

The threshold factor $\Gth$ is used for
gap extraction.
The algorithm for gap extraction is as follows:
\begin{itemize}
    \item For a given plunger and cutter gate voltage, we scan for the maximal non-local conductance value as a function of bias voltage and magnetic field. This provides a reference point for an extended state in our wires for the given junction configuration.
    
    \item We define the induced gap as the maximal bias voltage at which the conductance is below $\Gth = 0.05 \cdot G_\mathrm{max}$, where $G_\mathrm{max}$ is the maximal non-local conductance (see below). The induced gap is a function of the magnetic field.

    \item At the bias voltage and magnetic field at which we extract the maximal conductance, the device is thus necessarily above the induced gap as defined above. Quoting \RefTGPe:
    \begin{quote}
        ``For the disorder strengths expected in our devices, we take $\Gth$ equal to $\exp(-3) \approx 0.05$ times the maximal value $\mathrm{max}\left\{\GNL\right\}$ of the non-local conductance at bias voltages greater than the induced gap (scanning over all $B$ for each $V_{\mathrm{p}}$ for a given cutter configuration).''
    \end{quote}
\end{itemize}

Therefore, the protocol implemented in the code is consistent with our description in \RefTGPe~and this point in \RefLegga~is not correct.

\begin{figure}[ht!]
    \centering
    \includegraphics[width=\linewidth]{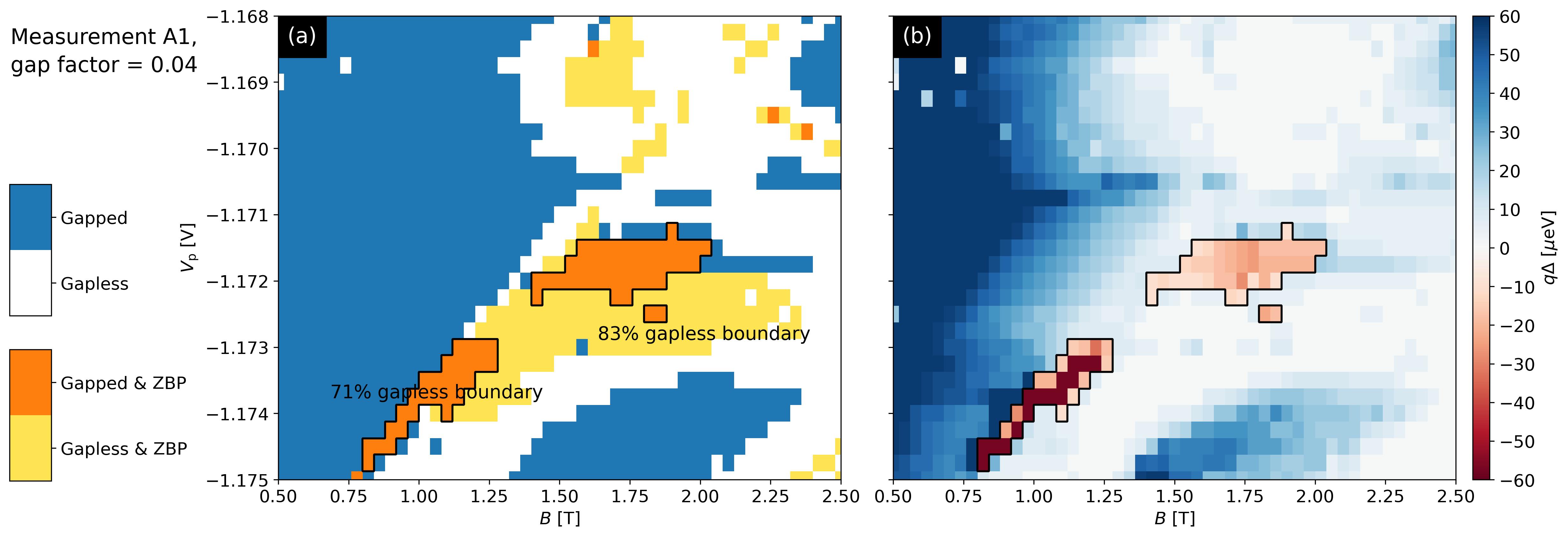}
    \includegraphics[width=\linewidth]{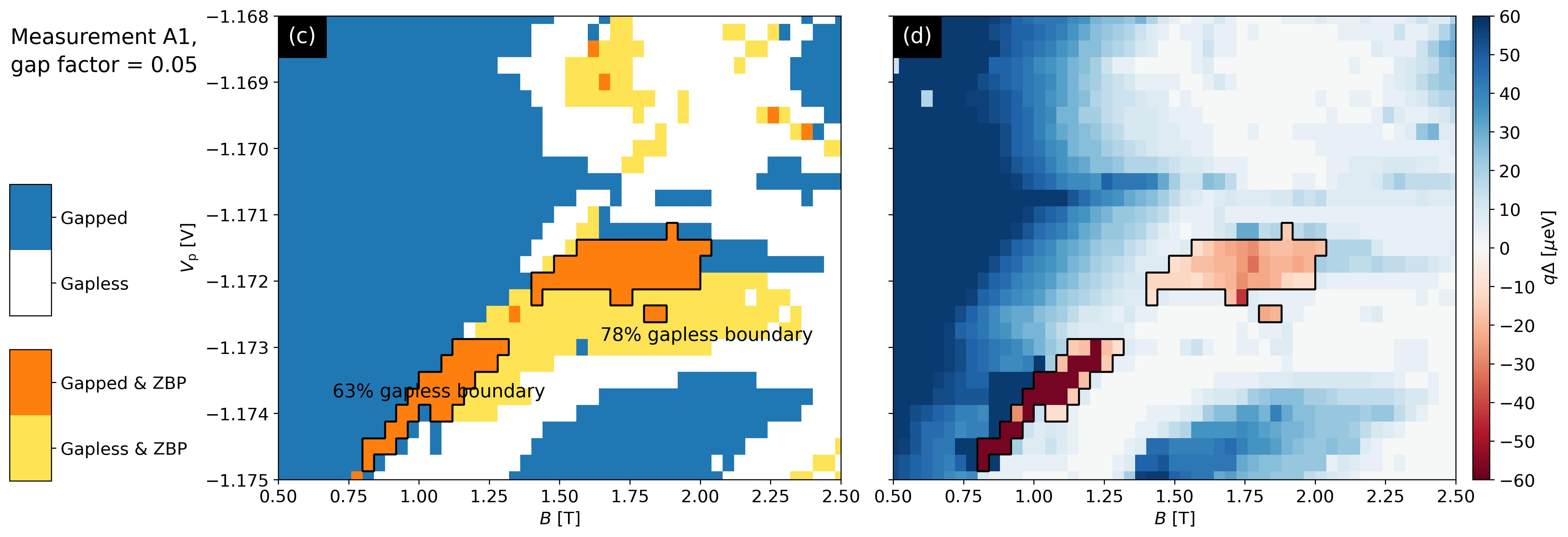}
    \includegraphics[width=\linewidth]{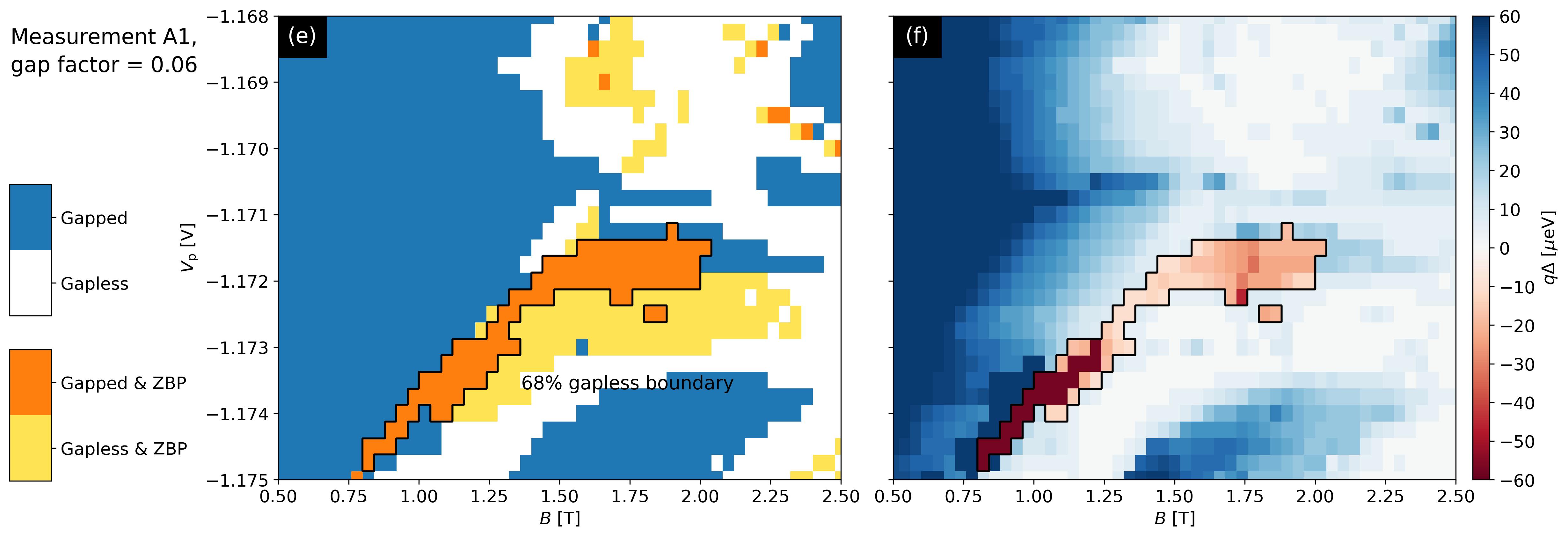}
    \caption{{\bf An example showing that a small change in the definition of $\Gth$ does not change the presence/absence of ROI2s.} As the conductance threshold parameter is varied from $0.04\cdot G_\mathrm{max}$ to 
    $0.06\cdot G_\mathrm{max}$
    for Measurement A1 data from \RefTGPe
    , only small variations in the the output of the TGP result, as expected.}
    \label{fig:cluster-threshold-sensitivity}
\end{figure}

\RefLegga~correctly points out the importance of the threshold $\Gth$ to the protocol:
\begin{quote} 
``The quantity $\Gth$ is therefore, arguably, the most important in the TGP since it determines whether a gap is reported by the
topological gap protocol.''
\end{quote}
For this reason, we studied the choice of threshold value in simulations.
The choice of threshold can affect the classification of individual pixels. See e.g. Fig.~\ref{fig:cluster-threshold-sensitivity} for an example of a sensitivity analysis as a function of threshold value. Small changes
in the threshold value of the conductance may lead to different subsets of the same cluster. 
While we have the freedom to choose $\Gth$ to maximize the accuracy of the protocol, it is nevertheless important that the output remains relatively robust against small changes in $\Gth$. Specifically, changes in $\Gth$ of less than $20\%$ result in only minor variations in the FDR (i.e. a few percent).

\RefLegga~then continues with the following statement.
\begin{quote} 
``It should be noted that this choice of 5\% highlights that the TGP has no general applicability.''
\end{quote}

This is not correct: as a statistical test, the protocol is generally applicable, and its hyperparameters are optimized according to the material stack and device design. It is usual for statistical models to have tunable hyperparameters, which is why we follow the standard practice of training and then cross-validating the model against independent datasets.

\subsection{2. Difference between analysis of simulated and experimental data}
\RefLegga~points out a minor difference between the analysis of experimental and simulated data. We agree.
The default parameter \lstinline{average_over_cutter} was set to \lstinline{True} for experimental datasets and to \lstinline{False} for simulated ones. The correct choice is \lstinline{average_over_cutter=True}. However, this difference does not lead to a statistically significant difference in the outcome of the TGP. Performing the analysis of experimental data with \lstinline{average_over_cutter=False} shows no significant difference in the outcomes presented in \RefTGPe. At the same time, applying the analysis with \lstinline{average_over_cutter=True} to simulated data yields the
outcome shown in \Cref{tab:TGP_FDR}. The FDR does not differ in any statistically significant way. Thus, this minor difference does not affect the reliability of the TGP.

\begin{table}[ht!]
\begin{center}
\begin{tabularx}{0.9\linewidth}{|c|c|c|c|c|}
\cline{1-5}
\shortstack[l]{\noalign{\vskip 1.0ex} Design, \\ stack} &
\shortstack[c]{$\sc$ \\ $[\scu]$} &
\shortstack[c]{TP \\ $\phantom{1}$} &
\shortstack[c]{FP \\ $\phantom{1}$} &
\shortstack[c]{FDR \\ $\phantom{1}$}
\\ [0.5ex]
\cline{1-5}
\multirow{3}{*}{SLG-$\beta$}
& 1.0 & 269 & 0 & $< 1.4$ \\ 
\cline{2-5}
& 2.7 & 57 & 0 & $< 6.3$ \\ 
\cline{2-5}
& 4.0 & 45 & 0 & $< 7.9$ \\
\cline{1-5}
\multirow{3}{*}{DLG-$\varepsilon$}
& 0.1 & 133 & 0 & $< 2.7$ \\ 
\cline{2-5}
& 1.0 & 115 & 0 & $< 3.2$ \\ 
\cline{2-5}
& 2.7 & 82 & 0 & $< 4.4$ \\ 
\cline{2-5}
& 4.0 & 79 & 1 & $< 6.8$ \\
\cline{1-5}
\end{tabularx}

\end{center}
\vskip -3mm
\caption{{\bf Simulated TGP statistics do not
change appreciably when \lstinline{average_over_cutter=True} as in experiments.}
Statistics of TGP results for simulated transport data from SLG-$\beta$ and DLG-$\varepsilon$ devices with \lstinline{average_over_cutter=True}.
}
\label{tab:TGP_FDR}
\end{table}

This table can be readily validated using the \lstinline{notebooks/yield_analysis.ipynb} notebook available in the code repository \cite{azure-quantum-tgp} and modifying \href{https://github.com/microsoft/azure-quantum-tgp/blob/d31e955de75b76b87d853ebc769a4f5f2220cbdb/notebooks/yield_analysis.py#L189}{this line in the code}. The code and package environment file for reproduction are provided.

\RefLegga~asserts that \RefTGPe~contains 
\begin{quote}
``the claim by Microsoft Quantum that
their simulations — however they were actually performed—
contain “no false positives”.''
\end{quote}
but \RefLegga~neglects to mention that the possibility of false positive TGP outcomes is explicitly mentioned in \RefTGPe. 
The text of \RefTGPe~states
\begin{quote}
    ``Since we found no false positives, the confidence interval for the FDR is between zero and the upper bound that we list in the rightmost column. 
    We find that if a device passes the TGP, there is a $< 8\%$ probability that the ROI$_2$ that it finds does {\rm not} contain a topological phase, provided that the simulated data is drawn from the same probability distribution as the data produced by real devices.
    For the DLG-$\varepsilon$ design, the probability is $< 6\%$.''
\end{quote}
This passage states that the probability of the occurrence of false positives is bounded by less than $8\%$. Therefore, the appearance of a single false positive due to the difference in TGP settings out of approximately 700 ROI2s does not undermine the validity of our claim that the probability of a TGP pass being a false positive is small.

\subsection{3. Variations in experimental data parameters}

\RefLegga~contains the claim \cite{Legg25a}:
\begin{quote}
    ``the TGP is not an unbiased test for topology, but instead, produces results that are dictated by measurement choices''
\end{quote}

We first note that this is a false dichotomy. The TGP is applied to a specific set of measurement data from a given device
(either experimental or simulated). Small changes in measurement choices lead to small
changes in the data. This, in turn,
can sometimes result in slight variations of the TGP outcome
(e.g.~\Cref{fig:cluster-threshold-sensitivity}),
but it can still be an unbiased test for topology. There is no contradiction. The precise shape of the identified region may change slightly if measurement choices are slightly changed. But this will not cause a statistically meaningful change in the probability of a false positive.
In the rest of this subsection, we re-iterate the reasons for the different parameter ranges in different measurements and show that the statistics of TGP outcomes are insensitive to these variations.

\subsubsection{Measurement Ranges in the TGP}
To understand the measurement ranges, we first recall that the TGP is composed of two stages. As we also describe in~\RefTGPe, stage 1 of the TGP is used as a coarse tool to identify promising clusters of correlated ZBPs, and consists of a scan over plunger voltage $\Vp$, in-plane magnetic field $\Bpara$, and cutter voltages $\Vc$ at zero bias.  
In TGP stage 1, we cover a $\sim 50$~mV range in $\Vp$ around the wire depletion voltage, which is determined in a separate measurement (see Appendix D1 of \RefTGPe), and $\Bpara$ in a $1-4$~T range. 
The exact ranges for $\Vp$ vary slightly, depending on device stability (see Sections IV.A.1 and IV.A.2 of \RefTGPe) and practical considerations such as the time the measurement would take.

We next move to describe the motivation for the ranges chosen for TGP stage 2, which is used to make a binary pass/fail statement. TGP stage 2 is performed around promising clusters identified in stage 1. The $\Vp$ ranges are typically chosen to extend several mV above and below the ZBP cluster identified in TGP stage 1. Typically this results in a $3-20$~mV range for $\Vp$, with a $100-\SI{250}{\micro\volt}$ resolution. For the $\Bpara$ range, we extend the field by several hundred mT on each side of the ZBP cluster, which results in a $\approx 2$~T range for $\Bpara$ with a $10-\SI{50}{mT}$ resolution. These values are chosen to ensure we capture enough of the gapless boundary of the ROI.
Here, too, there may be minor differences between measurements to account for device stability and to keep measurement times within reason ($\simeq 24$~hours).
For each TGP stage 2 run, the entirety of the resulting measured dataset is analyzed to determine whether the TGP passes or fails.

Similarly, the raw voltage bias range for TGP stage 2 is chosen to be large relative to the expected topological gap but small enough that the measurements do not take a prohibitively long time. In practice, this means the raw bias voltage range was chosen $> \SI{60}{\ueV}$ at positive and negative bias, with a $<\SI{10}{\micro\volt}$ resolution, where larger resolution was chosen for larger $\Vbias$ ranges.

As discussed in \RefTGPe,
measurement A1 was performed in a different cooldown from measurements A2 and A3. The disorder potential is expected to change from one cooldown to another, so it is natural for the ROI2 position change. Measurements A2 and A3 were performed a week apart within the same cooldown, during which time there was some device drift.
Consequently, the measurement ranges were different in these
three measurements of device A, which explains the ranges shown for this device in Fig.~2 of \RefLegga.

Summarizing, the sizes of stage 1
clusters determine the
measurement ranges in stage 2.
Larger stage 1 clusters will
lead to larger stage 2 data ranges.
Stage 2 measurement ranges also include additional margins around stage 1 clusters.
Taking larger margins
around stage 1 clusters enables a more precise ROI determination
in stage 2 but requires longer measurement times.

\subsubsection{Data Ranges in Simulations}
As explained in \RefTGPe,
we have tested our protocol in simulations that follow the same procedure of using the outcome of TGP stage 1 to determine the plunger gate voltage and field range of a stage 2 simulation. 
\begin{figure}[ht!]
    \centering
    \includegraphics[width=\linewidth]{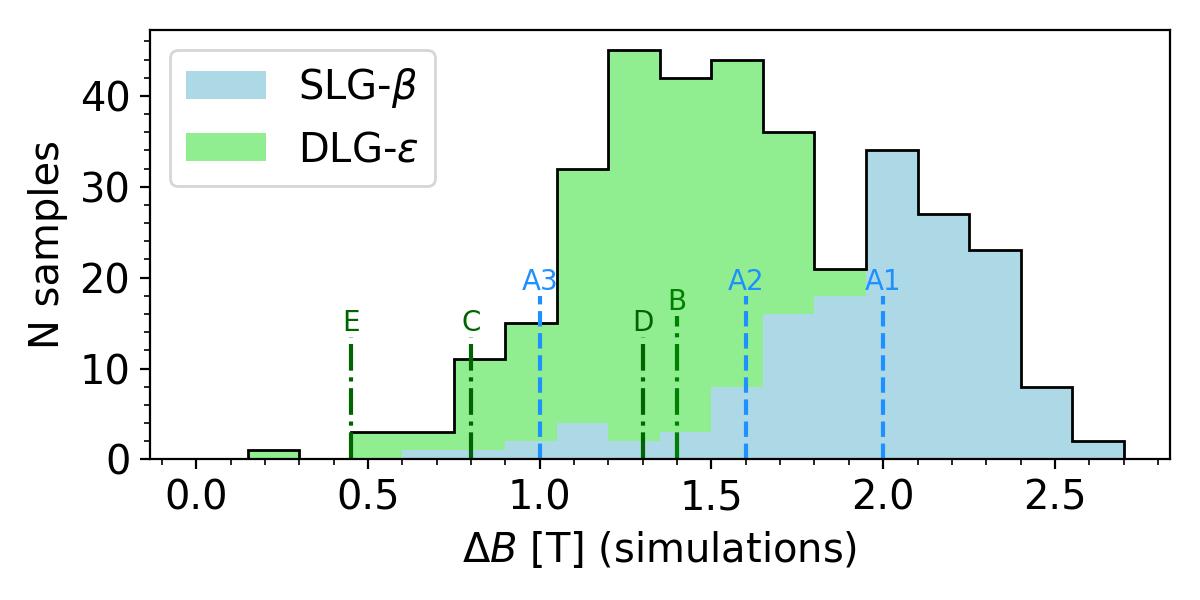}
    \caption{{\bf There is large variation in the stage 2 magnetic field range in simulations, as in experiments.} Distribution of magnetic field ranges $\Delta B$ over which TGP stage 2 was run in simulations. There is significant variation, determined by the stage 1 cluster sizes. This range is comparable to the range in the measured devices in \RefTGPe, which was highlighted in Fig.~2 of \RefLegga. We find a low FDR in these simulations.}
    \label{fig:range-distribution}
\end{figure}
The resulting distribution of the field ranges in stage 2 is shown in Fig.~\ref{fig:range-distribution}.
The ranges over which stage 2 was measured in experiment are well within this distribution.

\subsubsection{Stability of TGP
Statistics to Changes of the Magnetic Field Range}
We have ensured that the magnetic
field and voltage ranges
are large enough that we have a low FDR, as we now describe.
Using simulations that accurately model our devices (as determined by independent characterization measurements), we have optimized the hyperparameters of the protocol so that the TGP as implemented on devices has a low FDR.

For these simulations, we have shown that, as defined in \RefTGPe, the TGP yields a small ($<8\%$) FDR, which we consider acceptable.
This also means that we accept that in many cases a device will have a topological phase but still fail the TGP.

A low FDR is sufficient to justify
the choices of hyperparameters
of the TGP. However, we can make a stronger statement, which is that the
FDR does not change much
even if we reduce the magnetic field
range ``by hand.''
In \RefTGPe, we noted that:
\begin{quote}
``We do not have a single false positive in these data. 
This does not mean that the TGP perfectly identifies the topological region. \\...
When we restrict the magnetic field to $B \leqslant 2.5\,$T, we find one false positive ...''\\
\end{quote}
We expand on this point in
\Cref{tab:TGP_FDR_2.5T}, where we
show that the FDR remains low
even if we artificially restrict
the magnetic field range
to $B\leq \SI{2.5}{\tesla}$ \footnote{This table contains our result for the SLG-$\beta$ stack, where we have enough statistics. Most of the ROI2s are at high fields for the DLG-$\varepsilon$ stack.}.

\begin{table}[h!]
\begin{center}
\centering
\begin{tabularx}{0.9\linewidth}{|c|c|c|c|c|}
\cline{1-5}
\shortstack[l]{\noalign{\vskip 1.0ex} Design, \\ stack} &
\shortstack[c]{$\sc$ \\ $[\scu]$} &
\shortstack[c]{TP \\ $\phantom{1}$} &
\shortstack[c]{FP \\ $\phantom{1}$} &
\shortstack[c]{FDR \\ $\phantom{1}$}
\\ [0.5ex]
\cline{1-5}
\multirow{3}{*}{SLG-$\beta$,~$B\leq\SI{2.5}{\tesla}$}
& 1.0 & 207 & 1 & $< 2.6$ \\ 
\cline{2-5}
& 2.7 & 45 & 0 & $< 7.9$ \\ 
\cline{2-5}
& 4.0 & 39 & 0 & $< 9.0$ \\
\cline{1-5}
\end{tabularx}
\end{center}

\vskip -3mm
\caption{{\bf In simulations, the FDR has weak dependence
on the magnetic field range.}
Statistics of TGP results for simulated transport data from SLG-$\beta$ devices with  $B_{max} \leq \SI{2.5}{\tesla}$ (which is
$\SI{0.5}{\tesla}$ smaller than in \Cref{tab:TGP_FDR}). A comparison with 
\Cref{tab:TGP_FDR} shows that a change in
magnetic field range causes a very small change
in the FDR bound.}
\label{tab:TGP_FDR_2.5T}
\end{table}

\begin{figure}[t!]
    \centering
    \includegraphics[width=\linewidth]{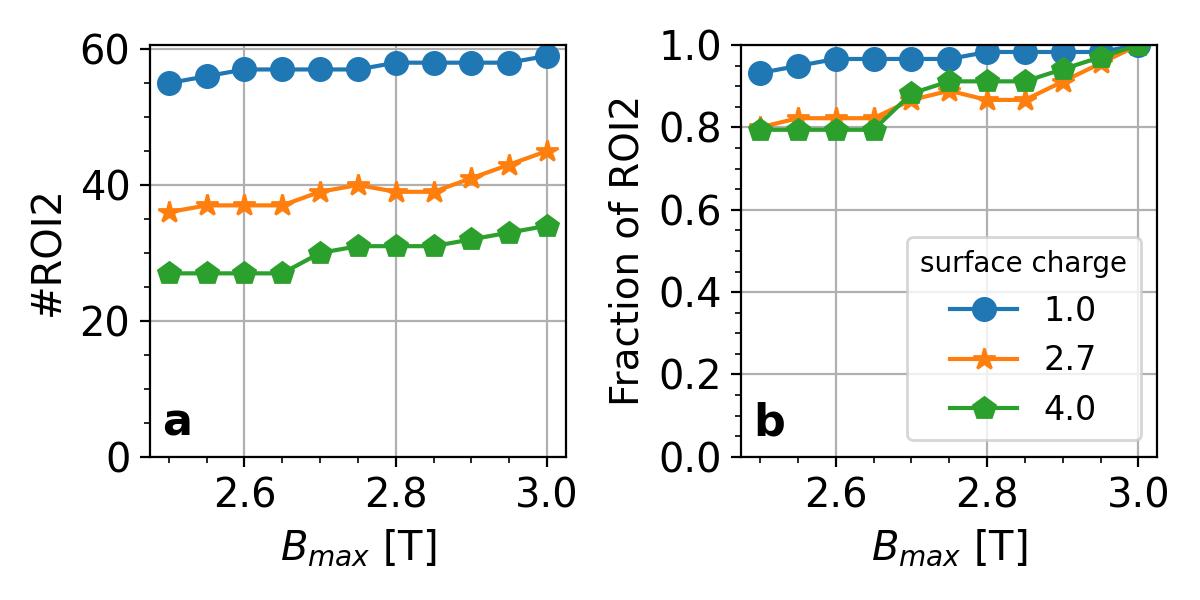}
    \caption{{\bf ROI2s are stable with
    respect to small changes in the magnetic field range, as tested in simulations.} We reduce the top end of the TGP Stage 2 range, $B_{max}$, ``by hand'' from $\SI{3}{\tesla}$ to $\SI{2.5}{\tesla}$. As we do this, we track the ROI2s that lie at least partially below $\SI{2}{\tesla}$. 
    (a) The number of ROI2s that overlap with the initial ROI2s found for $B_{max}=\SI{3}{\tesla}$ and lying entirely below $B=\SI{2}{\tesla}$. (b) The fraction of
    ROI2s that overlap with the initial ROI2s found for $B_{max}=\SI{3}{\tesla}$ and lying entirely below $B=\SI{2}{\tesla}$.
    We observe that more than $80\%$ of ROI2s 
    survive when we decrease the magnetic field range by $|\Delta B| = \SI{0.5}{\tesla}$.}
    \label{fig:field_range_sensitivity}
\end{figure}

\subsubsection{Robustness of Individual
ROI2s to Changes of the Magnetic Field Range}

We now address the sensitivity of the TGP to changes in magnetic field
range by tracking individual ROI2s.
As can be seen in \Cref{fig:field_range_sensitivity},
most ROI2s survive
as the field range is cropped
by hand.
Combined with the
analysis already presented in \RefTGPe, this shows that moderate changes in the magnetic field range (i.e. $\Delta B_{\rm{max}} \leq \SI{0.5}{\tesla}$) lead to small changes in TGP outcomes: less than $20\%$ of ROI2s are strongly modified, as shown in \Cref{fig:field_range_sensitivity}.
However, the FDR remains low, as explained in \Cref{tab:TGP_FDR_2.5T}, so changing the magnetic field range by $\SI{0.5}{\tesla}$ does not affect the reliability of the TGP, contrary to what is suggested in the comment.

\RefLegga~claims that
\begin{quote}
    ``Together with the magnetic field dependence, the TGP outcome can be therefore be selectively passed (or failed) by choosing a data range window that provides the desired result.''
\end{quote}

This is false.
It is not possible to selectively pass or fail the TGP because the analysis is performed after the data is collected. Additionally, it is impossible to predict what might occur outside the parameter range of the measurement.

\RefLegga~observes that
\begin{quote}
    ``the bias range determines the maximum possible reported gap''
\end{quote}
This point is correct. Precisely for this reason,
we have always taken a bias range larger
than the expected topological gap values.

\subsection{4. Definition of `topological'}

\RefLegga~claims that the requirement for the system to be counted as topological was changed when the TGP was tested.
It is important to emphasize that 
\RefLegga~refers to a change between {\it two different papers}, not within a single paper.
It is not surprising that the protocol and criterion evolved between the papers.
In \RefTGPe, we stated the criterion used in that paper
and tested the TGP against that criterion.
It is consistent with existing literature \cite{Akhmerov11}.

\RefLegga~states that 
\begin{quote}
``Perhaps not so surprisingly this weakened definition leads to 37.7$\%$ of all phase space being identified as ‘topological’.''
\end{quote}

\begin{figure}[ht!]
    \centering
    \includegraphics[width=\linewidth]{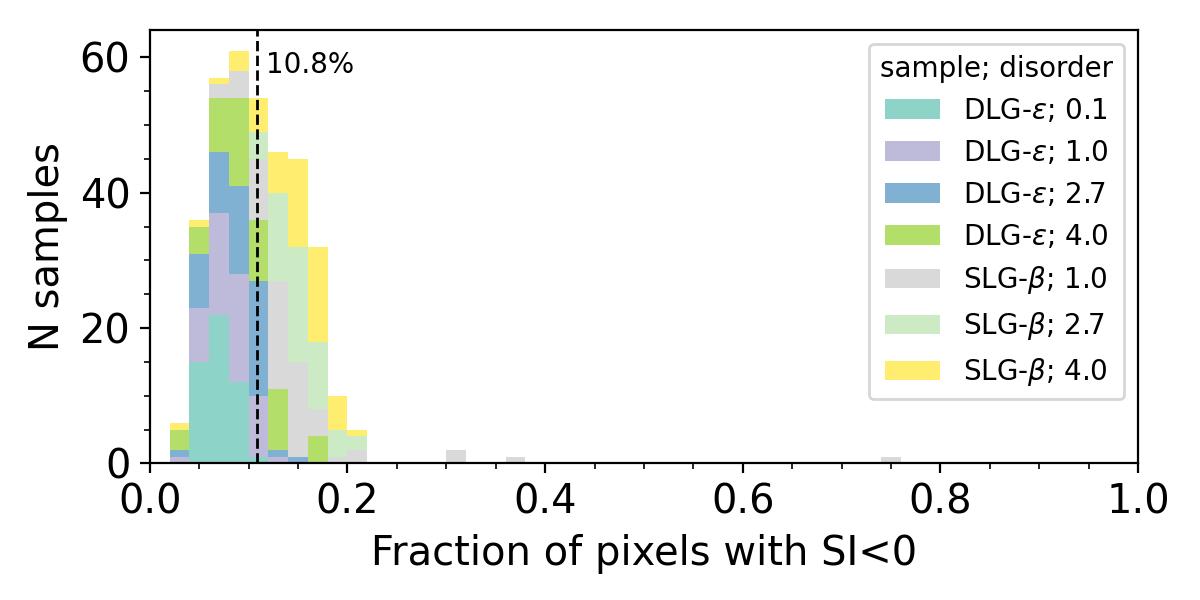}
    \caption{{\bf The topological phase is a small percentage of the total available phase space in simulations.} Distribution of fraction of pixels with negative scattering invariant in simulated data provided in Ref.~\cite{azure-quantum-tgp}.}
    \label{fig:topo_pixel_dist}
\end{figure}

This statement is incorrect. 
To see this, one can investigate the datasets publicly available in \lstinline{'data/simulated/yield/stage1'}. 
Inspecting the datasets, one can find the distribution of ``topological" vs ``trivial" pixel points in our data shown in Fig.~\ref{fig:topo_pixel_dist}. 

The median ratio of pixels classified as topological is 11$\%$ for all material stacks and disorder levels. The largest median ratio for any single materials stack and disorder
level occurs for SLG-$\beta$ at $\sc = 4.0 \times \scu$, where it is $15\%$.
Furthermore, our stage 1 scans of simulated devices cover $65\,$mV in voltage units, whereas the tuning range for the system is approximately $0.5\,$V (see e.g. Figs.~4 and 9 in \RefTGPe~for the plunger voltage ranges in, respectively,
simulations and measurements), and the rest of that
range is not expected to have regions of topological phase. Hence, the
topological phase actually occupies a
fraction of phase space that is 10 times
smaller than the median ratio that we
calculated above.

\RefLegga~observes that
\begin{quote}
``In other words if $\mathrm{det}(r) < 0$ is satisfied at either end of the nanowire, for any cutter, and for a single pixel, then the whole region counts as a ‘true positive’... Instead, this altered definition allows mixed trivial and ‘topological’ regions to be treated as a single phase, inflating the apparent success of the protocol.''
\end{quote}

This deduction is incorrect:
in fact, the success of the protocol is
not inflated. The goal
of the TGP is to determine if a device
supports the topological phase and to find its location within the large phase space of a device (which would otherwise be unmanageable).
If we want to use the device for topological qubits, then we need to tune it to the
right magnetic field and plunger gate voltage
(possibly for many wires), and our results show
us that the TGP enables this. Naturally, we would then find the best operating point for
MZM-based qubits with further measurements, such as those in \RefZMPR. Some parts of the region identified by
the TGP may be poor operating points for
parity measurements, either because they are not in the topological phase
or because the MZMs simply have suboptimal coupling to the leads. But the TGP will have succeeded at accomplishing its goal.

This section of \RefLegga~then concludes:
\begin{quote}
    ``Overall this redefinition of ‘topological’ means that the claimed
‘true positives’ in the simulations performed for \RefTGPe~can be unrelated
to MBSs.''
\end{quote}
This assertion is not supported by
any of the arguments in \RefLegga. Specifically, \RefLegga~does not provide a statistical probability for this occurrence.  

The definition of topological is subtle in
finite-sized systems with disorder, but
\RefLegga~does not address this subtlety,
thereby undermining its conclusion.
Most definitions of topological phases depend on the scaling with system size; for instance, topological degeneracy implies energy splittings that decay exponentially $\sim e^{-L/\xi}$. (See Ref.~\onlinecite{Day25} for a recent discussion of the finite-size
scaling of a topological index.)
However, an experimental measurement deals with a single device, and it is important to have
a criterion that can be applied to a
single device. $\mathrm{Det}(r)=0$ is a natural boundary for the topological phase, since it signifies the closing of the transport gap. Once we identify the topological region (i.e.\ SOI), we characterize the magnitude of the topological gap via non-local conductance.    

Moreover, there is an even stronger
argument that true positives correspond to the
topological phase. A true positive
must have at least a single pixel with stable zero bias peaks, finite topo gap and negative invariant in the region of interest -- i.e. it must pass the TGP and also have negative invariant.
While the scattering invariant might not
capture all of the physics, this is further
addressed by the presence of stable
zero bias peaks at both ends of the wire
and the re-opening of the bulk gap.
The criterion for a true positive is
extremely stringent. And our results show that $>92\%$ of TGP passes have regions that also have negative invariant.

\RefLegga~has a figure titled ``Inclusion of `missing topological pixels....' There are no missing pixels in Fig. 32 in our paper. It appears that this sentence is referring to the overlay of $\mathrm{SI}<0$ hatching which the author has added in panel (ii). Most of this figure has $\mathrm{SI}<0$. There are three points worth making here. (1) We have not systematically studied the probability of false negatives, but it is not surprising that they occur.
(2) This shows an important way in
which the TGP is a higher bar than
the scattering invariant:
topological regions with small
gap will not pass the TGP even though they may, technically,
be topological. It is plausible that only useful
topological regions pass the TGP, which is essential for the tuneup of
more complex devices.
(3) Of course, the stage 2 scan covers a small fraction of the device's parameter space, so this does not imply that $\mathrm{SI}<0$ over the entire phase diagram. As noted above, it is less than $11\%$ of simulated devices' phase space.

\section{Technical response to Ref. 4}

We begin this section by noting that \RefLeggb\ does not challenge the main claim
of \RefZMPR: 
\begin{quote}
    ``... we observed a flux-dependent bimodal RTS in the quantum capacitance, which we interpret as switches of the parity of a fermionic state in the wire.'' 
\end{quote}

Instead of questioning the main claim,
\RefLeggb~argues that the device
tuneup procedure outlined in
Subsection S4.3 of the online
Supplementary Information for
\RefZMPR~does not clearly show
that there is a superconducting gap in the wire.

In fact, the minimal requirement for any explanation of the main result of
\RefZMPR~is the presence of a fermionic state in the wire
that is (a) close to zero energy, (b) connects the two ends of the quantum-dot interferometer, (c) is separated in energy from other states in the wire, and (d) is proximitized with similar weight of particle and hole components.
The hypothesis of a gapless system
in \RefLeggb~is ruled out by the analysis in Sec. S2.7, where the extra pair of MZMs can represent an additional extended mode in the wire. The size of the observed quantum capacitance signal is only consistent with additional modes that are well-separated (by at least $k_B T$) from the fermionic zero mode.
Thus, the claim of a gapless system \cite{Legg25b}
is not consistent with
the results of our rf measurements. 
Since \RefLeggb~arrives at its claims through
an analysis of the TGP data in
Subsection S4.3 of the online
Supplementary Information for
\RefZMPR~,
we will address it point-by-point below
even though, as just mentioned, its primary contention
is contradicted
by our parity measurements.

\paragraph{TGP can report a region to be gapped or gapless.}
We have addressed the sensitivity to data ranges in the response to \RefLegga. 
Since the same arguments apply, we won't repeat them here.

\paragraph{Correctness of TGP outcomes.}
The peer review file contains the authors' response to a referee's query about whether parity measurements were performed elsewhere in the phase diagram: ``In each measurement, this was the only region passing the TGP within the explored gate voltage and magnetic field range.'' \RefLeggb~places considerable emphasis on this
correspondence and alleges that it is incorrect:
\begin{quote}
``This statement is incorrect. As shown in Fig.~3(ii), applying the TGP as implemented in \RefTGPe~but now with all regions identified by the TGP shown, Device~B exhibits multiple regions that are identified by the TGP. In particular, in Fig.~3 there is a region ($V_{\rm p}\approx-1.6725$~V, $B\approx 2.8$~T) that is identified by the TGP and where no `parity readout' data was presented (a further third region in this device is present for other cutters and also passes the TGP). The only reason this region was not identified in the TGP phase cartoon shown in \RefTGPe~was because the input to the TGP plotting code caused  only the largest identified region to be highlighted.''

\end{quote}

The allegation made in
\RefLeggb~is false. In the correspondence quoted above, we state that, within the parameter range explored in the parity measurement
(namely, fixed $B = \SI{2.0}{T}$, and $V_p$ range corresponding to the full TGP range), there was only one region that passed the TGP. In other words these were the
only pixels passing the TGP in this slice, which is a correct statement.

We note that the published version of the TGP data in Subsection S4.3 of the Supp. Info. for Ref.~\cite{Aghaee25} has a bug causing an incorrect cropping of the bias range (i.e. it is not symmetric around zero bias, with 1 extra pixel at negative bias), as pointed out in \RefLeggb. This resulted in an artificial lowering of extracted gap values. Correction of this bug leads to small changes in the extracted gap values (in more than 96\% of pixels, the change is smaller than $\SI{5}{\micro\electronvolt}$). However, due to the thresholding method used in the TGP, those changes could lead to a change in the status of a pixel (gapless to gapped or visa versa), which could lead, in turn, into a change of cluster overlap, and the subsequent (dis-)appearance of ROI2s. In measurements A1 and A2, the bug fix leads only to minor changes in the shape of ROI2s, on a level of few pixels. For measurement B, the bug fix results in the 
appearance of a new SOI2 for one cutter and a consequent increase in the size of an ROI2. This by itself, however, does not necessitate any change to the statement in our correspondence with the referee.

\paragraph{Presence of a gap based on conductance data.}

\RefLeggb~states

\begin{quote}
``The inconsistent and incorrect TGP outcomes for the data of Ref.~\citenum{Aghaee25} further evinces that the TGP is not a reliable diagnostic tool for the existence of an SC gap in the nanowire. As such, we instead investigate the actual conductance data for each device. It should be emphasized that these data were not presented in Ref.~\citenum{Aghaee25}, even for the specific parameter regions that passed the TGP. Only TGP phase `cartoons' are presented in the supplemental material of Ref.~\citenum{Aghaee25}. These conductance data reveal that the regions where `parity readout' occur are actually highly disordered and do not present features consistent with a well-defined superconducting gap.''
\end{quote}

These claims are factually incorrect.

(a) This commentary quoted above is preoccupied with the positioning of the TGP output in Subsection S4.3 of the Supplementary
Information for \RefZMPR, juxtaposed against the conductance data residing in the Zenodo repository. Regardless of
the location within the Supplementary
Information or the Zenodo
repository, there is no evidence
that the system is highly-disordered.

\begin{figure*}
    \centering
    \includegraphics[width=\linewidth]{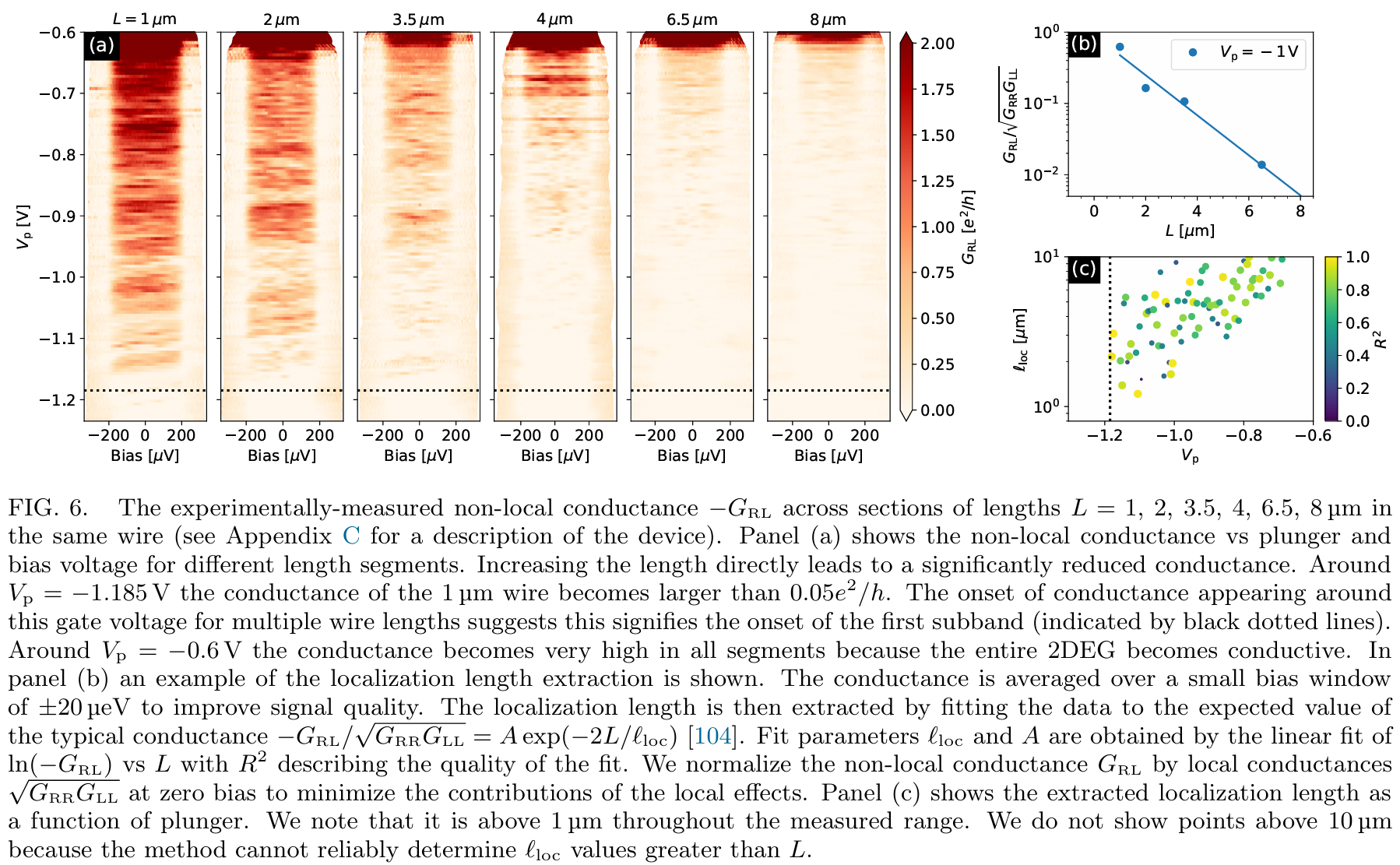}
    \caption{{\bf Our devices have low disorder.} We have reproduced Fig.~6 from~\cite{Aghaee23} here. It shows significant non-local conductance in wires that are longer than $\SI{3}{\micro\meter}$. Exponential fits of the dependence on wire length show that the localization length in the lowest sub-band is $>\SI{1}{\micro\meter}$.}
    \label{fig:localization}
\end{figure*}

(b) We have characterized the disorder level in our proximitized nanowires in \RefTGPe. The normal-state localization length in these wires
(which is the integrated measure of disorder in these devices)  is on the order of a micron or more in the single subband regime,
as may be seen in \Cref{fig:localization}, in
which we have reproduced a plot from \RefTGPe.
This demonstrates that our process yields devices with low disorder.

(c) The logic behind the
claim in \RefLeggb~that
the data is not consistent with
``a well-defined superconducting gap"
is the following:

\begin{quote}
``Starting with the local conductance in Figs.~4, 5, 6, it can be seen that there are a very large number of low energy states. Although the case for all devices, this is particularly striking for Device A2, cutter 2 [Fig.~2(f)], where there is strong finite local conductance, $G_{\rm RR}$, for almost the full range of bias voltages, indicating an abundance of low-energy states and there is also no clear SC gap features. It is also interesting to note that in all cases there is a sizable difference in magnitude of the local signal on each side of the nanowire.''
\end{quote}

The claims above are based on misconceptions and imprecise language regarding the relationship between conductance spectroscopy and the density of states in the wire. Specifically, the comment refers to device A2, cutter 2, and claims ``strong finite local conductance, $G_{RR}$, for almost the full range of bias voltages, indicating an abundance of low-energy states.'' In the limit of weak coupling to the transport lead, the local conductance is directly proportional to the local density of states. However, in the strong coupling regime, this relationship changes. In this limit, the presence of subgap conductance does not necessarily indicate a finite density of states; this phenomenon is known as Andreev enhancement of the local conductance (see Fig. 7 in \cite{Blonder82}). A discrete state strongly coupled to a transport lead can produce a broad feature in voltage bias. We believe this is the case here, as low bias conductance level reaches values above 2$e^2/h$. Therefore, no features of the local conductance require or even strongly suggest an interpretation involving ``an abundance of low-energy states,'' contrary to the claim in \RefLeggb.

Secondly, the comment claims that the conductance data show ``significant nonlocal conductance at low-bias consistent with a gapless system.'' However, the non-local conductance is not a direct measure of the density of states in the wire. Instead, it measures the transmission through the wire, which we use to define the transport gap in a finite-size wire. The non-local conductance generally has components that are both symmetric and antisymmetric in bias. The symmetric component may arise due to the energy dependence of the local matrix elements or the leakage of an Andreev state from one side to the other in a finite-size wire \cite{Danon20,kurilovich24}. This does not imply a breaking of particle-hole symmetry. Importantly, both the symmetric and antisymmetric components of the non-local conductance are proportional to the transmission through the wire. Therefore, an extraction of the transport gap requires thresholding of the non-local conductance, which needs to be done separately for the symmetric and antisymmetric components. We note that there is a practical advantage in choosing the antisymmetric component of the non-local conductance for gap extraction because it is less affected by voltage divider effects~\cite{Martinez21}. Conversely, the presence of a strong resonance in the junction might significantly impact the symmetric component, necessitating careful subtraction of voltage divider effects.

Third, the comment claims the conductance data show evidence of particle-hole symmetry breaking: ``In particular, considerable antisymmetric components of local conductance and symmetric components of nonlocal conductance are extremely clear [...]" Particle-hole symmetry does not forbid anti-symmetric components of
the local conductance nor symmetric components of the non-local conductance in a 3-terminal configuration. Symmetry relations have been extensively studied in the literature \cite{Danon20, Menard20, Maiani22}. Once again, the assertion regarding particle-hole symmetry breaking is unsupported by any of the arguments in \RefLeggb.

\section{Discussion}
\label{sec:alternatives}

As we have shown, the
remarks in Refs.~\onlinecite{Legg25a,Legg25b} lack substantive relevance to the
main claims of Refs.~\onlinecite{Aghaee23,Aghaee25}.
Instead, they reveal
a major disconnect between two different approaches to identifying a one-dimensional topological superconductor.

The first approach is to find a “smoking gun” signature which is consistent with the topological phase but is qualitatively inconsistent with the trivial phase. Implicitly, the signature of the topological phase would be sufficiently universal as to be discernible from the trivial phase even if the device were treated as a black box.

The second approach is statistical. It is based on a recognition that data taken from a topological superconducting device can (in principle) also be obtained from a device that does not have a topological phase but, instead, has “accidental” fine-tuned zero-energy states in some parameter regimes of the trivial phase. However, if the data 
contain features that are sufficiently
robust as many parameters are varied, then it is very unlikely to be due
to accidental fine-tuned zero-energy states. The probability that the device has a topological phase can be quantitatively estimated provided some features of the device -- such as the disorder level – are taken into account. Naturally, there will be edge cases which could be mis-classified, and the probability of these can be estimated. This is the false discovery rate (FDR).

Our papers \cite{Aghaee23,Aghaee25} have taken the second approach. However, Refs.~\cite{Legg25a, Legg25b} would be meaningful only in reference 
to a paper that had taken the first approach (sensitivity to protocol parameters, statements about “broad applicability,” measurement ranges). This explains why \RefLegga~does not contradict the main result of \RefTGPe:
it does not dispute that we have InAs-Al devices that pass the TGP.
It similarly explains why
\RefLeggb~does not contradict the main result of \RefZMPR:
it does not dispute that we observed a flux-dependent bimodal RTS in the quantum capacitance.
Instead, Refs.~\cite{Legg25a, Legg25b} are concerned with ways in which the second approach differs from the first approach. But the two approaches are necessarily different.

The main criteria of success for TGP is a low false discovery rate. Refs.~\onlinecite{Legg25a,Legg25b}
have not found any evidence
that the FDR is significantly higher
than previously
estimated \cite{Aghaee23}
or challenged the main claims of
Refs.~\onlinecite{Aghaee23,Aghaee25}.

\begin{acknowledgments}
We thank Anton Akhmerov and Michael Wimmer for fruitful discussions.
\end{acknowledgments}

\vspace{1mm}
\textbf{Correspondence and requests for materials} should be addressed to Chetan Nayak~(cnayak@microsoft.com).

\vspace{1cm}
$^\dagger${Morteza Aghaee, Zulfi Alam, Mariusz Andrzejczuk, Andrey E. Antipov, Mikhail Astafev, Amin Barzegar, Bela Bauer, Jonathan Becker, Umesh Kumar Bhaskar, Alex Bocharov, Srini Boddapati, David Bohn, Jouri Bommer, Leo Bourdet, Samuel Boutin, Benjamin J. Chapman, Sohail Chatoor, Anna Wulff Christensen, Patrick Codd, William S. Cole, Paul Cooper, Fabiano Corsetti, Ajuan Cui, Andreas Ekefjärd, Saeed Fallahi, Luca Galletti, Geoff Gardner, Deshan Govender, Flavio Griggio, Ruben Grigoryan, Sebastian Grijalva, Sergei Gronin, Jan Gukelberger, Marzie Hamdast, Esben Bork Hansen, Sebastian Heedt, Samantha Ho, Laurens Holgaard, Kevin Van Hoogdalem, Jinnapat Indrapiromkul, Henrik Ingerslev, Lovro Ivancevic, Thomas Jensen, Jaspreet Jhoja, Jeffrey Jones, Konstantin V. Kalashnikov, Ray Kallaher, Rachpon Kalra, Farhad Karimi, Torsten Karzig, Maren Elisabeth Kloster, Christina Knapp, Jonne Koski, Pasi Kostamo, Tom Laeven, Gijs de Lange, Thorvald Larsen, Jason Lee, Kyunghoon Lee, Grant Leum, Kongyi Li, Tyler Lindemann, Matthew Looij, Marijn Lucas, Roman Lutchyn, Morten Hannibal Madsen, Nash Madulid, Michael Manfra, Signe Brynold Markussen, Esteban Martinez, Marco Mattila, Robert McNeil, Ryan V. Mishmash, Gopakumar Mohandas, Christian Mollgaard, Michiel de Moor, Trevor Morgan, George Moussa, Chetan Nayak, William Hvidtfelt Padkær Nielsen, Jens Hedegaard Nielsen, Mike Nystrom, Eoin O'Farrell, Keita Otani, Karl Petersson, Luca Petit, Dima Pikulin, Mohana Rajpalke, Alejandro Alcaraz Ramirez, Katrine Rasmussen, David Razmadze, Yuan Ren, Ken Reneris, Ivan A. Sadovskyy, Lauri Sainiemi, Juan Carlos Estrada Saldaña, Irene Sanlorenzo, Emma Schmidgall, Cristina Sfiligoj, Sarat Sinha, Thomas Soerensen, Patrick Sohr, Tomaš Stankevič, Lieuwe Stek, Eric Stuppard, Henri Suominen, Judith Suter, Sam Teicher, Nivetha Thiyagarajah, Raj Tholapi, Mason Thomas, Emily Toomey, Josh Tracy, Michelle Turley, Shivendra Upadhyay, Ivan Urban, Dmitrii V. Viazmitinov, Dominik Vogel, John Watson, Alex Webster, Joseph Weston, Georg W. Winkler, David J. Van Woerkom, Brian Paquelet Wütz, Chung Kai Yang, Emrah Yucelen, Jesús Herranz Zamorano, Roland Zeisel, Guoji Zheng, Justin Zilke}

\bibliography{response}

\end{document}